\documentclass[a4paper,12pt]{article}

\usepackage{geometry,slashed}

\usepackage
{
   amsmath,amssymb,graphicx,tabularx
}
\usepackage{soul}
\usepackage[usenames,dvipsnames]{xcolor} 
\usepackage{color}
\usepackage[verbose,colorlinks=true,naturalnames=true,linkcolor=black,urlcolor=black, citecolor=black]{hyperref}
\usepackage{caption,subcaption}
\usepackage{authblk}
\usepackage{epstopdf}
\usepackage{enumerate}
\usepackage{multirow}
\usepackage{siunitx}
\usepackage{float}
\usepackage{booktabs}
\usepackage{cite}
\usepackage{pdflscape} 
\usepackage{listings}
\usepackage{diagbox} 
\DeclareMathAlphabet{\pazocal}{OMS}{zplm}{m}{n}
\graphicspath{ {./}}
\usepackage{url} 
\usepackage{varwidth}
\usepackage{comment}
\usepackage[parfill]{parskip}

\newcommand{\madgraph}{\textsc{MadGraph5\_aMC@NLO}}
\newcommand{\mcatnlo}{\textsc{MC@NLO}}
\newcommand{\madspin}{\textsc{MadSpin}}
\newcommand{\pythia}{\textsc{Pythia8}}
\newcommand{\rivet}{\textsc{Rivet}}

\numberwithin{equation}{section}

\title{Matrix Element Corrections in top quark decays for the $t\overline{t}W^{\pm}$ process}

\author{Rikkert Frederix\thanks{rikkert.frederix@fysik.lu.se} }
\author{Leif Gellersen\thanks{leif.gellersen@fysik.lu.se, now employed at Axis Communications} }
\author{Jasmina Nasufi\thanks{jasmina.nasufi@fysik.lu.se} }

\affil{\small Division of Particle and Nuclear Physics, Department of Physics, Lund University, Box 118, SE-221 00 Lund, Sweden}

\begin{document}

\maketitle

\begin{abstract}
We present a method that allows enabling Matrix Element Corrections (MECs) in \pythia{} with \mcatnlo{} matching, without incurring double counting. MECs are an interesting feature that may contribute to the accuracy of theoretical predictions, alongside matching and merging. We directly compare our method to a specific choice of settings in \pythia{}, which can remove double-counting for MECs in certain processes. We show results by taking the $t\overline{t}W$ process as an example. This choice allows us to study the impact of decay MECs in the $2\rm{SS}\ell$ and $3\ell$ final states. We find that jet-related observables receive these corrections unevenly throughout the phase space. They can contribute up to $\pm 6\%$ in certain regions.    
\end{abstract}

\section{Introduction}
\label{intro}

In the era of the High-Luminosity LHC, the accuracy of theoretical predictions must keep up with experimental advances. Among the several approaches to achieve this goal, it is imperative to strive for better predictions by incorporating higher order perturbative corrections. Currently, NLO QCD and EW corrections, as well as matching to parton showers at NLO QCD accuracy are completely automated in mainstream event and parton shower generators (see e.g.\ refs.\ \cite{Alwall:2014hca,Frederix:2018nkq,Bierlich:2022pfr,Sherpa:2019gpd,Bahr:2008pv,Bellm:2015jjp}). An interesting feature in \pythia~\cite{Bierlich:2022pfr}, that can additionally contribute at this level of accuracy, are the so-called Matrix Element Corrections ~\cite{Bengtsson:1986hr,Bengtsson:1986et,Norrbin:2000uu,Cabouat:2017rzi}. We will refer to them as MECs in the following.\\
MECs correct the usual splitting kernel in the DGLAP evolution by a ratio factor of the exact matrix element over the approximate matrix element generated by the shower, thus potentially improving the baseline performance of parton showers. In the absence of higher-order perturbative calculations, MECs are a computationally efficient way that may enhance the accuracy of theoretical predictions and are therefore of phenomenological interest. In \pythia{} MECs are turned on by default, unless one is interested in \mcatnlo ~\cite{Frixione:2002ik,Frixione:2003ei} style matching. In this case, according to the official recommendation of the latest Pythia manual, all MEC functionalities must be turned off by the user to avoid potential double-counting. This is unfortunate, since MECs can e.g.\ correct resonance decays and promote them to approximate NLO QCD accuracy, which should give a better description of the radiation pattern of jet-related observables. The first goal of this paper, is to present a general method that remedies the seeming incompatibility between MECs and the \mcatnlo{} matching scheme. We note, that this problem has been the topic of a recent paper~\cite{Frixione:2023hwz}, which appeared while our work was in progress. In this paper, Frixione et al. suggest that the double counting may be avoided by selecting the appropriate settings in the current \pythia{} implementation. We directly compare both approaches by presenting predictions for the $t\overline{t}W^{\pm}$ process as an example.\\
This choice serves the second goal of the paper, which is to study the impact of decay MECs on the $t\overline{t}W^{\pm}$ process in the multi-lepton final states at the LHC. This process has been measured by both ATLAS~\cite{ATLAS:2015qtq,ATLAS:2016wgc,ATLAS:2019nvo,ATLAS:2019fwo,ATLAS:2023xay,ATLAS:2023gon} and CMS~\cite{CMS:2015uvn,CMS:2017ugv,CMS:2022tkv} in the $2\rm{SS}\ell$ and $3\ell$ final states. Normalization and shape differences relative to theoretical predictions have been observed since the beginning of measurements. Despite significant improvements on the theory front, these tensions have not been fully addressed yet (see e.g. ref. \cite{ATLAS:2023gon}). Thus it is important to keep pushing the frontier and simultaneously explore all available options, such as MECs.\\
We give a brief overview of the preceding research in calculating higher order corrections for $t\overline{t}W^{\pm}$. At the production level, fixed order predictions with NLO QCD+EW \cite{Frixione:2015zaa} and subleading EW \cite{Dror:2015nkp} effects are available. The complete-NLO corrections have been calculated in ref.\ \cite{Frederix:2017wme}, and have been matched to soft-gluon resummation to reach NLO+NNLL accuracy in ref.\ \cite{Broggio:2019ewu,Kulesza:2020nfh}. Earlier resummation work includes refs.\ \cite{Li:2014ula,Broggio:2016zgg,Kulesza:2018tqz}. Recent papers have presented updated predictions that include NNLO QCD corrections in ref.\ \cite{Buonocore:2023ljm}
and resummed and expanded predictions up to aN$^{3}$LO QCD with NLO EW corrections included in ref.\ \cite{Kidonakis:2023jpj}. A more realistic description of the process takes into account decays to the resonant particles. Refs.\ \cite{Campbell:2012dh,Bevilacqua:2020pzy,Bevilacqua:2021tzp} perform the calculation in the narrow-width approximation (NWA), whereas other studies go beyond the NWA and include full off-shell effects with NLO QCD \cite{Bevilacqua:2020pzy,Denner:2020hgg}, NLO QCD with an extra light jet \cite{Bi:2023ucp}, NLO QCD with subleading EW \cite{Bevilacqua:2021tzp} and complete-NLO corrections \cite{Denner:2021hqi}. These fixed order predictions describe the decays and spin correlations at NLO accuracy. At the cost of downgrading the accuracy of these latter aspects to LO, or ignoring them altogether, it is possible to venture beyond fixed order and improve the description of the final state radiation pattern. By using the \textsc{Powheg}~\cite{Nason:2004rx,Frixione:2007vw,Alioli:2010xd} and \mcatnlo~\cite{Frixione:2002ik,Frixione:2003ei} matching formalisms, parton shower matched predictions at NLO QCD accuracy \cite{Garzelli:2012bn,Maltoni:2014zpa,Maltoni:2015ena} and including subleading EW effects \cite{Frederix:2020jzp,FebresCordero:2021kcc} have been computed. Finally, in addition to matching, it is possible to enhance predictions by merging processes with increasing jet multiplicities. The \textsc{FxFx} merging formalism~\cite{Frederix:2012ps} has been employed in ref.\ \cite{vonBuddenbrock:2020ter} and later on improved in ref.\ \cite{Frederix:2021agh} to provide predictions for $t\overline{t}W^{\pm}$ up to two jets at NLO in QCD matched to parton showers.\\
The structure of this paper is organized as follows. We discuss MECs and our new method in section \ref{explain_MEC}. We define the computational setup of our calculation and results in sections \ref{setup} and \ref{results}, and conclude in section \ref{finish}.
%
%
%
%
%
%
%
\section{Matrix Element Corrections and Matching}
\label{explain_MEC}
%
\begin{figure}[t!!]
\centering
\includegraphics[scale=0.32]{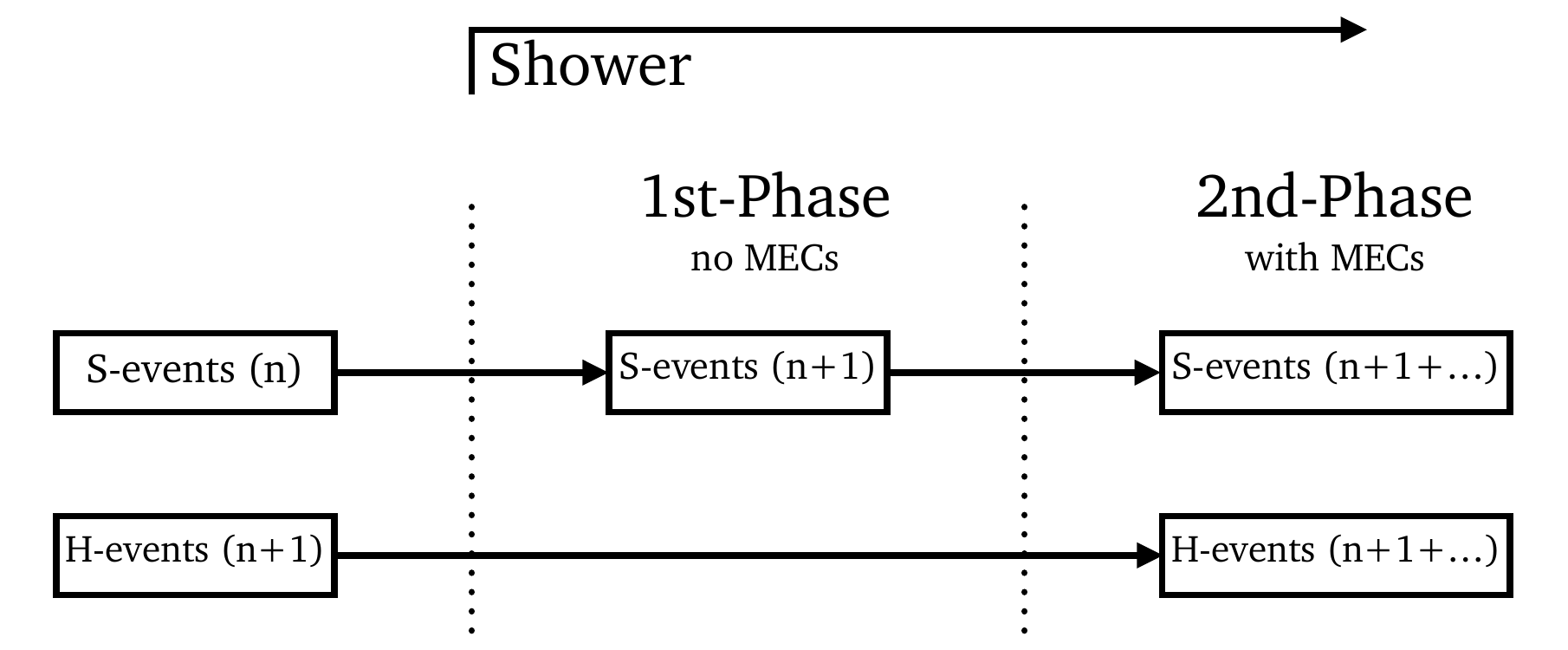}
\caption{Schematic representation of the $2$-step method to incorporate MECs.}
\label{fig0}
\end{figure}
In this section we elaborate on the incompatibility between MECs and \mcatnlo{} and propose a solution to remedy it. In addition, we also briefly summarize the alternative proposal by ref.\ \cite{Frixione:2023hwz}. \\
We take the $pp\rightarrow t\overline{t}W+X$ process as an example. \madgraph~\cite{Alwall:2014hca,Frederix:2018nkq} can generate hard matrix elements for $t\overline{t}W$ production at NLO in QCD and encode the relevant information in $\mathbb S$- and $\mathbb H$-type events. Hard $\mathbb H$-events contain resolved real emission kinematics and standard $\mathbb S$-events describe Born-like kinematics. If we pass this on to a parton shower algorithm, it can happen that shower emissions from $\mathbb S$-type events generate the same contribution as what is already included in the real-emission matrix elements. This kind of double contribution is accounted for and taken care of by the \mcatnlo{} matching formalism. \\
Turning on MECs in \pythia{} without proper care changes the emission spectrum, and can therefore introduce additional double counting that is not foreseen in the current implementation of \mcatnlo.  We begin by recognizing that this double-counting can occur only for the first emission of $\mathbb S$-type events. To avoid triggering it, we propose to enable MECs \textit{after} the first emission, see fig.\ \ref{fig0}. This leads us to divide the parton shower sequence in two phases, which we term the $2$-step approach:
\begin{itemize}
\item The first phase: We distinguish between $\mathbb S$-type and $\mathbb H$-type events, and allow for only one shower emission of $\mathbb S$-type events. This can be achieved by using \texttt{UserHooks} in \pythia{} on LHE files which contain information on the number of partons in the Born state. The $\mathbb S$-events are updated with a new shower scale according to refs.\ \cite{Bierlich:2022pfr,Brooks:2019xso} and extended with an additional parton, if present. The updated $\mathbb S$-events and the original $\mathbb H$-events are then written out into a new LHE file. Importantly, MECs are turned off during this phase. Furthermore, no QED radiation is allowed and emissions are treated in the global recoil scheme, as required by the \mcatnlo{} implementation in \madgraph.
\item The second phase: The $\mathbb S$-type and $\mathbb H$-type events from the first phase are showered as usual. Since the matching already takes place in the first phase, we may turn on MECs and QED effects. We are also free to choose a local recoil scheme.
\end{itemize} 
Heavy resonances are decayed after the first phase, before restarting the shower. Our implementation is valid for a $p_T$-ordered shower, such as the simple shower in \pythia. It is currently implemented in an external user file, and it not part of the default \pythia{} package.\\
We have cross-checked our method in two different ways. First we have compared the updated shower scale we calculated with the internal shower scale from \pythia{} for all relevant splittings at double digit precision. Secondly, we performed a ``0-th order check". The idea behind it, is that the $2$-step shower should be equivalent to the undivided shower, if MECs and the QED shower are turned off. We have tested this for some events, by comparing shower scales and particle kinematics after each shower emission at double digit precision. Additionally, we have also tested it at the level of physical observables. We generated sufficient statistics for $t\overline{t}W^{\pm}$ and run both the $2$-step shower and the undivided shower. This comparison yielded distributions which are in agreement within the Monte Carlo (MC) uncertainties.\\
We note some subtle aspects of the implementation of the $2$-step approach, which may affect enabling certain settings in \pythia. Generally, if the shower setting requires a different treatment of shower emissions and extra emissions recorded in the LHE file (present in $\mathbb H$-events or in updated $\mathbb S$-events after the first phase), this might impact the final results. For example, the rapidity ordering procedure is impacted by the $2$-step approach, since it currently orders shower emissions within the same shower sequence, but it ignores extra emissions listed in the LHE file. Similarly, azimuthal asymmetries in the shower are also impacted. We disable these settings during the validation procedure we describe above and for the default setup of results, described in section \ref{setup}.\\
The approach proposed in ref.\ \cite{Frixione:2023hwz}, resolves the issue by simply disabling \textit{some} FSR MECs. One can distinguish between \textit{exact} and \textit{equivalent} MECs, implemented in different and independent settings in \pythia. To avoid the double-counting paradigm, they suggest turning off approximate MECs
\begin{itemize}
\item[] \texttt{TimeShower:MEcorrections = on}
\item[] \texttt{TimeShower:MEextended = off}~.
\end{itemize} 
Caution is advised when it comes to applying this, as it does not guarantee foregoing double-counting in general. But it is valid in the case of $t\overline{t}$, and we assume that that validity also extends to  $t\overline{t}W^{\pm}$ in the context of our study. For further details please see the original paper. 
%
%
%
%
%
%
\section{Setup}
\label{setup}
We compute predictions for $pp \rightarrow t \overline{t}W^{\pm}$  with decays into the $3\ell$ and $2\rm{SS}\ell$ final states at  the LHC, with $\sqrt{s}=13$ TeV. Events are generated at the production level with NLO QCD corrections at $\mathcal{O}(\alpha_S^2\alpha^2)$ using the \madgraph{} software  package. The decays of heavy resonances are executed by \madspin~\cite{Artoisenet:2012st} in the full spin mode. To generate physical events, we match to the simple dipole shower in \pythia{} (version 8.310) following the \mcatnlo{}  matching prescription. The final output is ran through a \rivet~\cite{Bierlich:2019rhm} analysis to generate the final results.\\
The main purpose is to investigate the impact of MECs on the $t\overline{t}W^{\pm}$ process. Specifically, we are interested in corrections to the decay products of the top-quarks. To this end, we enable MECs only in the final-state shower and disable other default \pythia{} options
\begin{itemize}
\item[] \texttt{SpaceShower:MEcorrections = off} ,\qquad \texttt{TimeShower:MEcorrections = on} ,
\item[] \texttt{TimeShower:MEafterFirst = off}   ,\qquad\hspace{0.3cm} \texttt{TimeShower:MEextended = off}~.
\end{itemize}
We note that \texttt{MEextended} \textit{must} be turned off in the approach suggested in ref.\ \cite{Frixione:2023hwz}, as discussed earlier. The option \texttt{MEafterFirst} controls MECs after the first emission correction. Strictly speaking, it doesn't have to be turned off for any consistency reasons. Nevertheless, we keep it turned off for our default predictions and report on its impact on results in the discussion. \\ 
We define three setups that shall be compared against one another:
\begin{itemize}
\item noMEC: In this setup no MECs are included.
\item 1-step MEC: In this setup we include MECs according to the approach proposed in ref.\ \cite{Frixione:2023hwz}. This prescription is briefly discussed in section \ref{explain_MEC}.
\item 2-step MEC: In this setup we include MECs according to the method we propose in section \ref{explain_MEC}. 
\end{itemize}  
In the following we document the various settings and parameters employed in our calculation. The beams are parameterized by the NLO PDF sets: NNPDF3.1~\cite{NNPDF:2017mvq} with $\alpha_s(m_Z)=0.118$ (ID:303400) in \madgraph{} and NNPDF3.1 QCD+LUXQED with $\alpha_s(m_Z)=0.118$ (ID:324900) in \pythia. The matrix elements are generated in the $5$ flavor scheme and we do not include any off-diagonal CKM-matrix contributions. We use the same set of SM inputs:
\begin{equation}
  \begin{tabular}{l l l}
    $M_t = 173.34 \; {\rm GeV}$, &  $M_Z = 91.1876\; {\rm GeV}$, &  $M_\tau = 1.777\; {\rm GeV}$,  \\
    $\Gamma_t = 1.49150 \; {\rm GeV}$, & $\Gamma_Z = 2.4414\; {\rm GeV}$, & $\Gamma_W = 2.0476\; {\rm GeV}$, \\
    $\alpha_{EW} = 1/132.232$, & $G_\mu = 1.16639 \times 10^{-5}\; {\rm GeV}^{-2}$, & $(\Rightarrow M_W = 80.385\; {\rm GeV})$.
  \end{tabular}
  \label{eq:input_par}
\end{equation}
as in ref.\ \cite{Frederix:2021agh}. The value used for the mass of the $W$-gauge boson is calculated from the electroweak coupling $\alpha_{EW}$ and the Fermi constant $G_{\mu}$. \\
In \madspin{} we decay (anti-)tops as $t\rightarrow Wb$ and arrange for different decays of the $W$-gauge bosons, depending on the final state signature. Specifically, for $3\ell$, only leptonic decays are permitted $W\rightarrow \ell \nu_{\ell}$ into all three lepton flavors. In the case of $2\rm{SS}\ell$, we differentiate based on the dominant charge. For example, for $2\rm{SS}\ell^+$, $W^+$-gauge bosons are decayed  leptonically, whereas $W^-$-gauge bosons are decayed into partons or $\tau^-$-leptons. Similarly for $2\rm{SS}\ell^-$. Tau leptons are decayed inclusively in \pythia.\\
We use the default dynamic scale setting in \madgraph{} for the  renormalization $\mu_R$ and factorization $\mu_F$ scale parameters, with 
\begin{eqnarray}
\mu_R=\mu_F=\mu=H_T/2 = \sum_{ i \in \{t,\overline{t},W^{\pm},(j)\}} m_{T,i}/2~~,
\end{eqnarray}
In \pythia{} we employ the default option of setting the starting shower scale equal to the scale of the hard events $\mu$. We enable QCD and QED showers in both initial- and final-states and allow for hadronization. MPI, detector-level effects and tagging efficiencies are not considered. We do not include theoretical scale uncertainties or consider FxFx merging either, since they are not expected to impact MEC effects.\\
The fiducial regions of $3\ell$ or $2\rm{SS}\ell$, with $\ell = \{e,\mu\}$, define the requirements on the number and charge of the leptons, which must also pass the fiducial cuts
\begin{equation}
|\eta(\ell)| <2.5 \qquad p_T(\ell)< 10 ~ \rm{GeV}~~.
\end{equation} 
Jets are formed by clustering hadrons and photons using the anti-$k_T$ jet algorithm with $R=0.4$. We distinguish $b$-jets from light jets, where the former is defined as a jet that contains at least one $B$-hadron and light-jets may contain other hadrons and photons. Long lived B-hadrons are kept stable. We request at least two such $b$-jets that pass the following cuts
\begin{equation}
|\eta(j)| <2.5 \qquad p_T(j)< 25 ~ \rm{GeV}
\end{equation} 
and are inclusive in the light jet spectrum.
%
%
%
%
%
%
%
\section{Results}
\label{results}
%
%
\begin{figure}[t]
\centering
\includegraphics[scale=0.49]{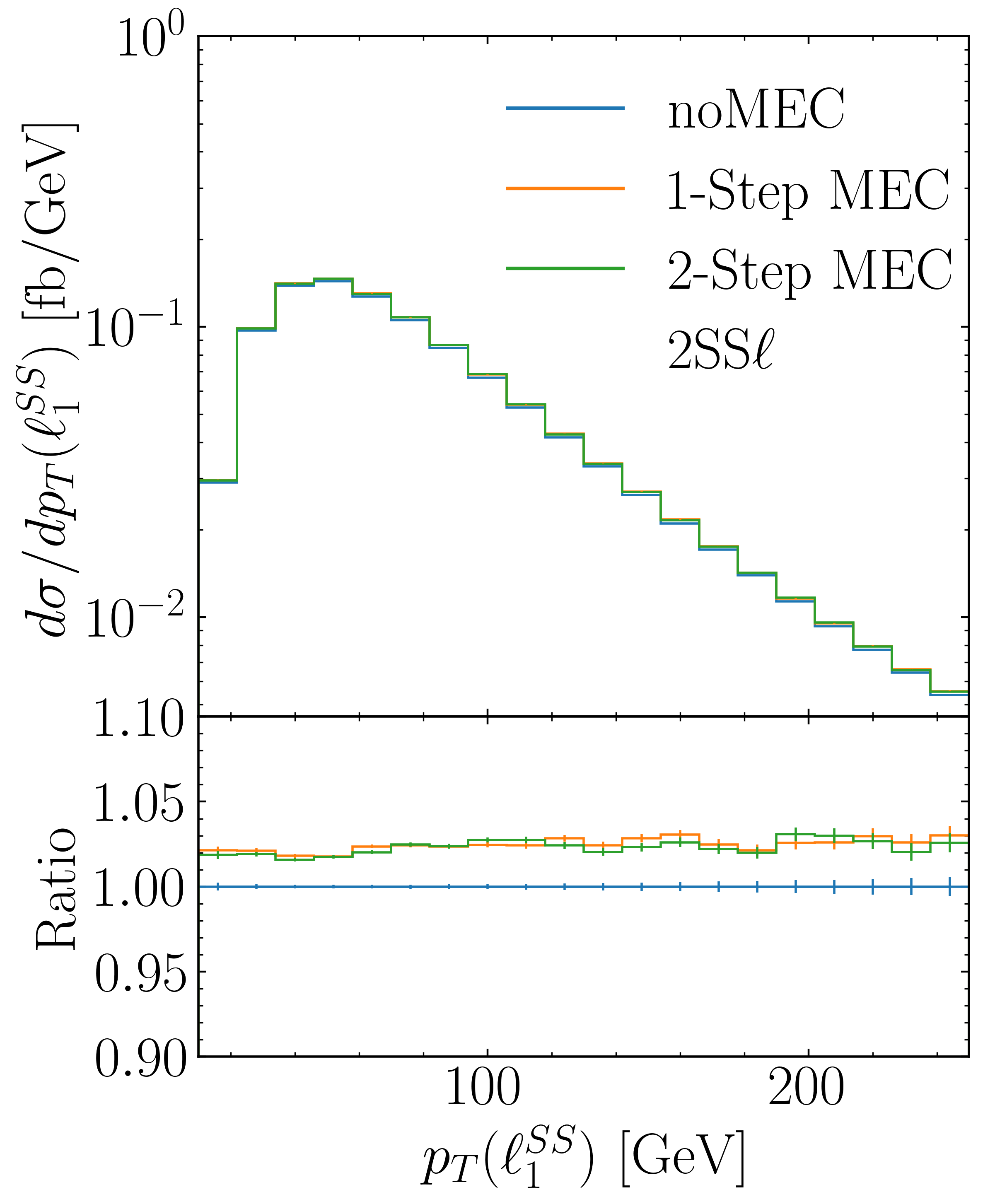}
\includegraphics[scale=0.49]{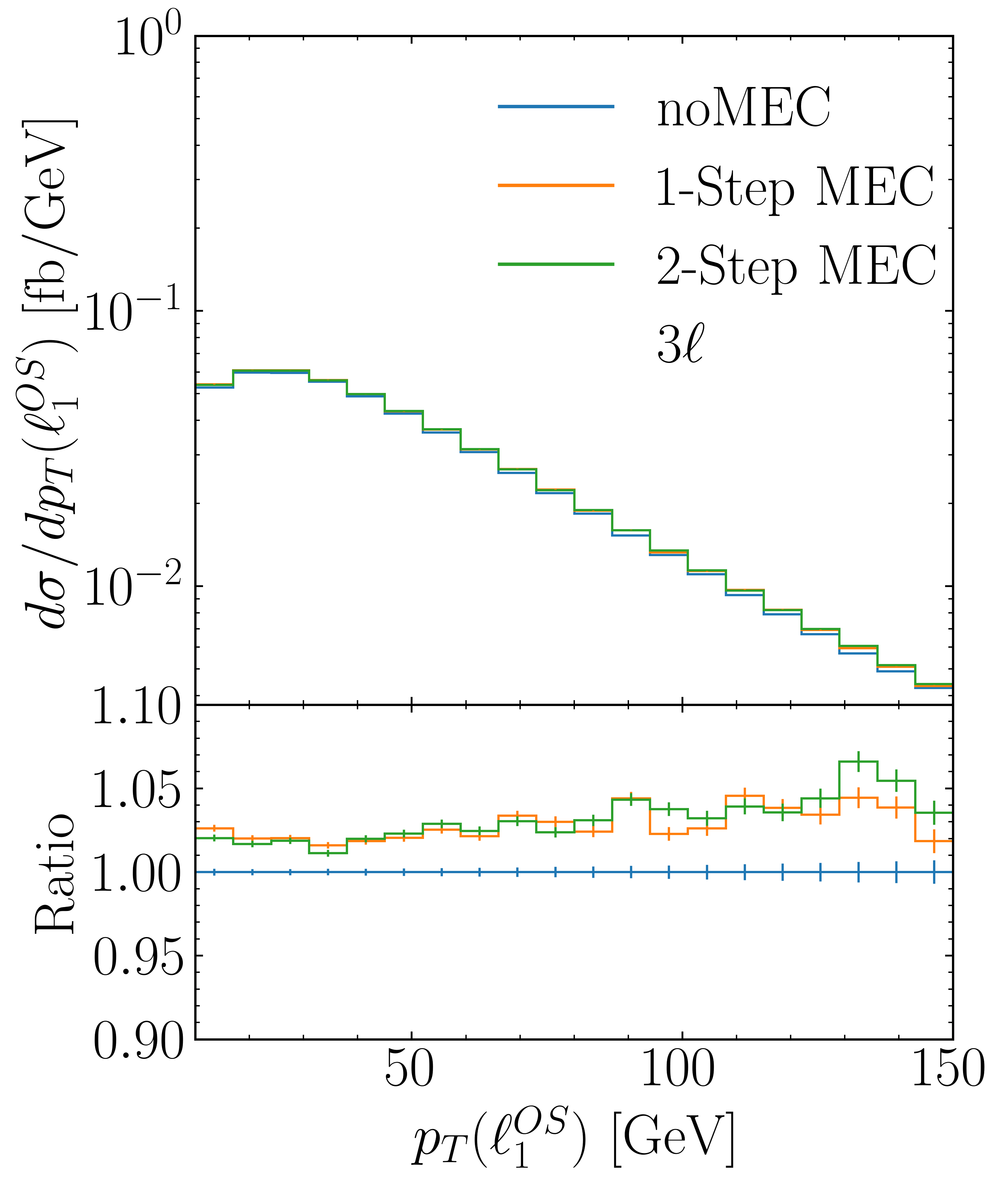}
\caption{The left plot shows the transverse momentum of the hardest same-sign lepton in the $2\rm{SS}\ell$ final state. The right plot shows the transverse momentum of the hardest opposite-sign lepton in the $3\ell$ final state. The upper panel contains absolute predictions for the setup without MECs and two setups with MECs included. The lower panel contains the ratio of all predictions to noMEC.}
\label{fig1}
\end{figure}
\begin{figure}[t]
\centering
\includegraphics[scale=0.49]{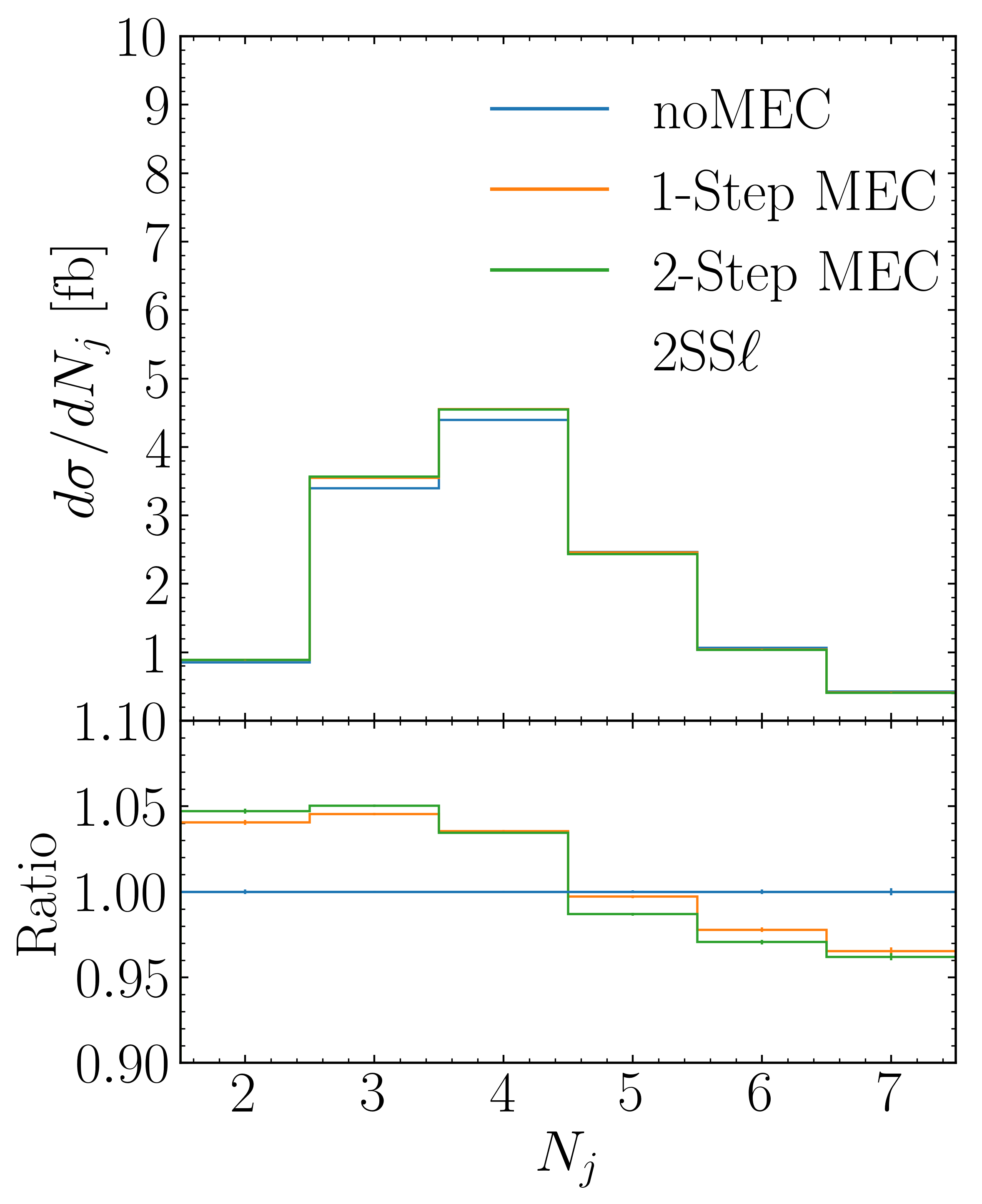}
\includegraphics[scale=0.49]{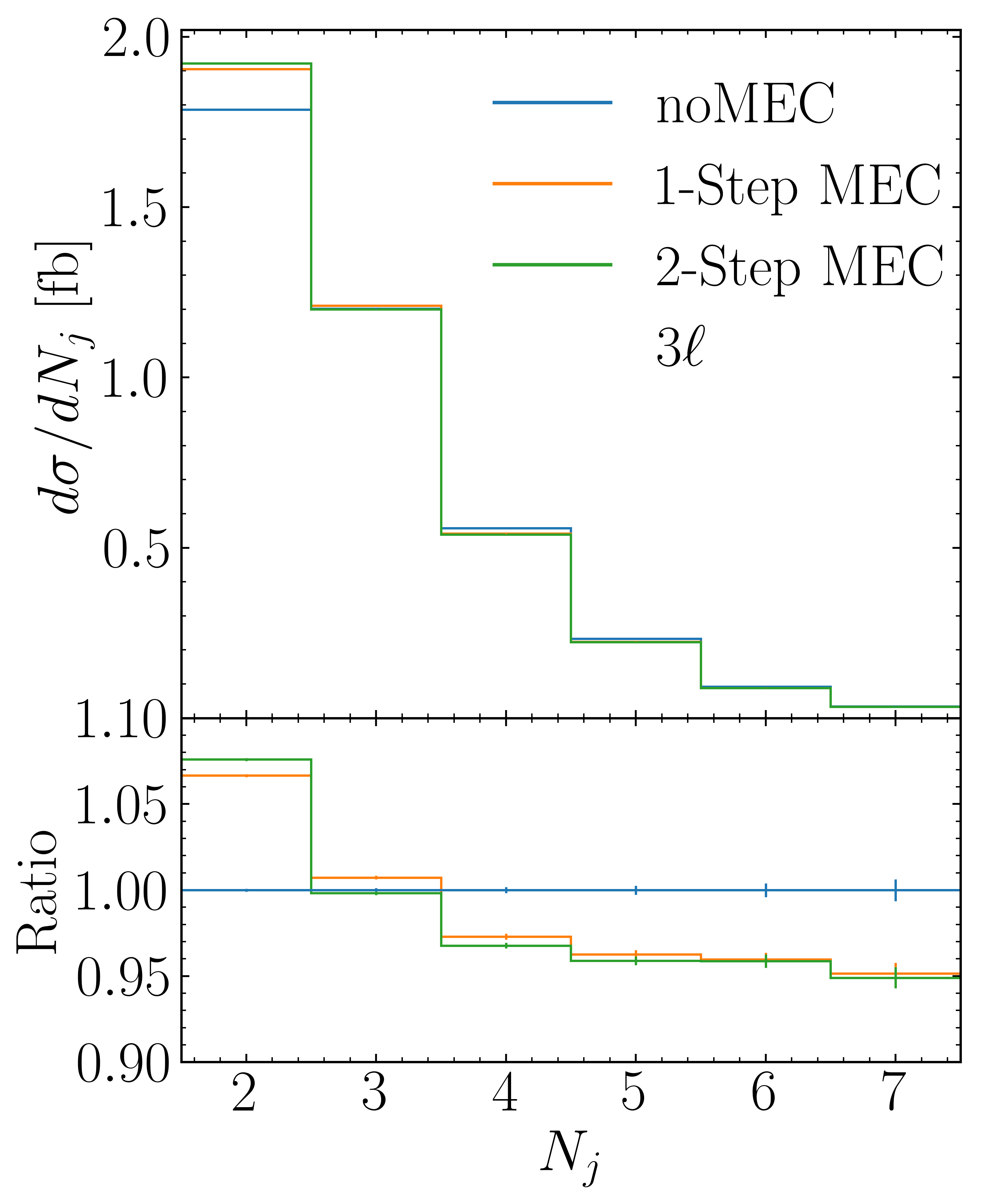}
\caption{The left plot shows the number of all jets in the $2\rm{SS}\ell$ final state. The right plot shows the number of all jets in the $3\ell$ final state. The upper panel contains absolute predictions for the setup without MECs and two setups with MECs included. The lower panel contains the ratio of all predictions to noMEC.}
\label{fig2}
\end{figure}
\begin{figure}[t]
\centering
\includegraphics[scale=0.49]{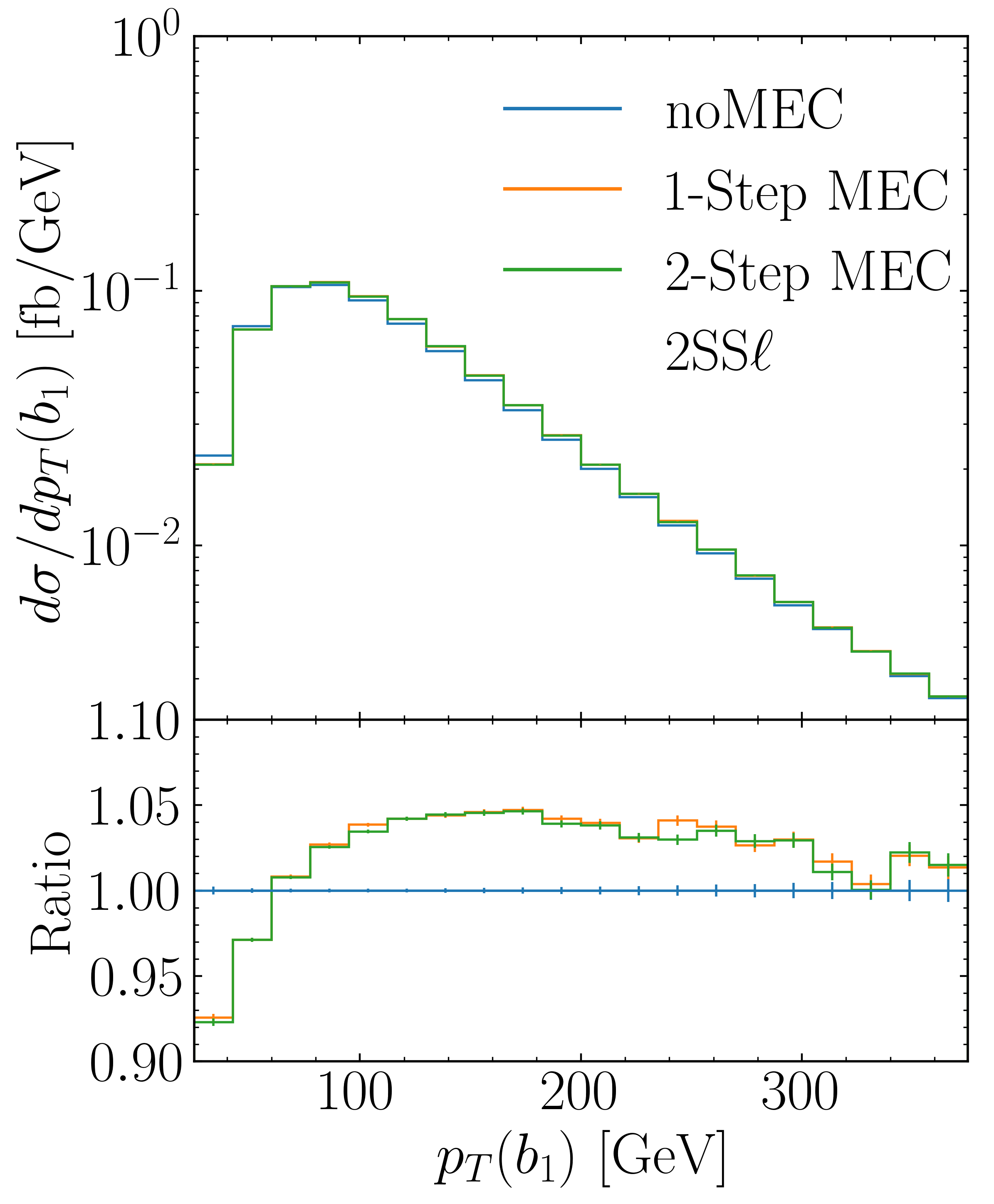}
\includegraphics[scale=0.49]{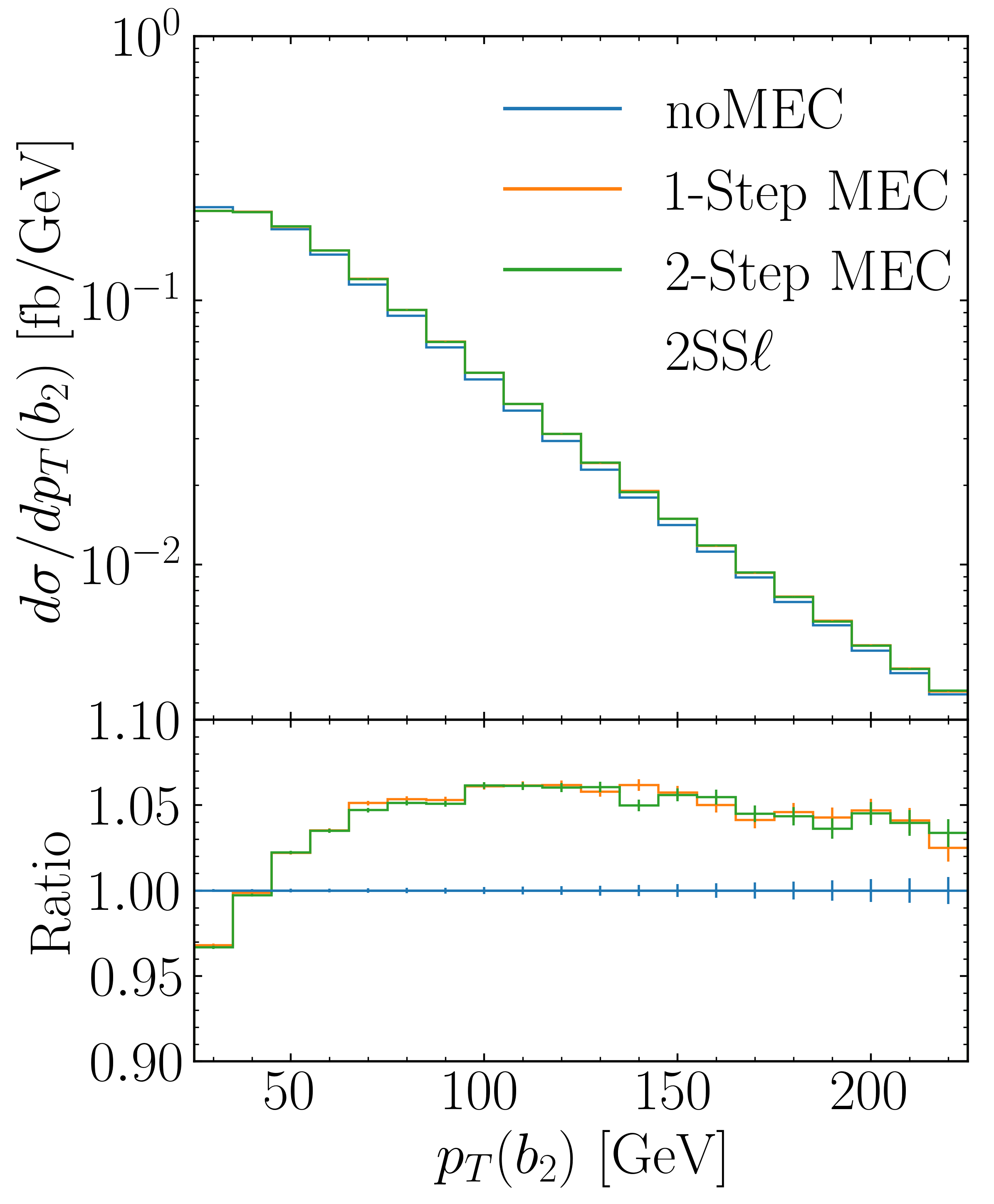}
\caption{The left plot shows the transverse momentum of the hardest $b$-jet in the $2\rm{SS}\ell$ final state. The right plot shows the transverse momentum of the second hardest $b$-jet in the $2\rm{SS}\ell$ final state. The upper panel contains absolute predictions for the setup without MECs and two setups with MECs included. The lower panel contains the ratio of all predictions to noMEC.}
\label{fig3}
\end{figure}
\begin{figure}[t]
\centering
\includegraphics[scale=0.49]{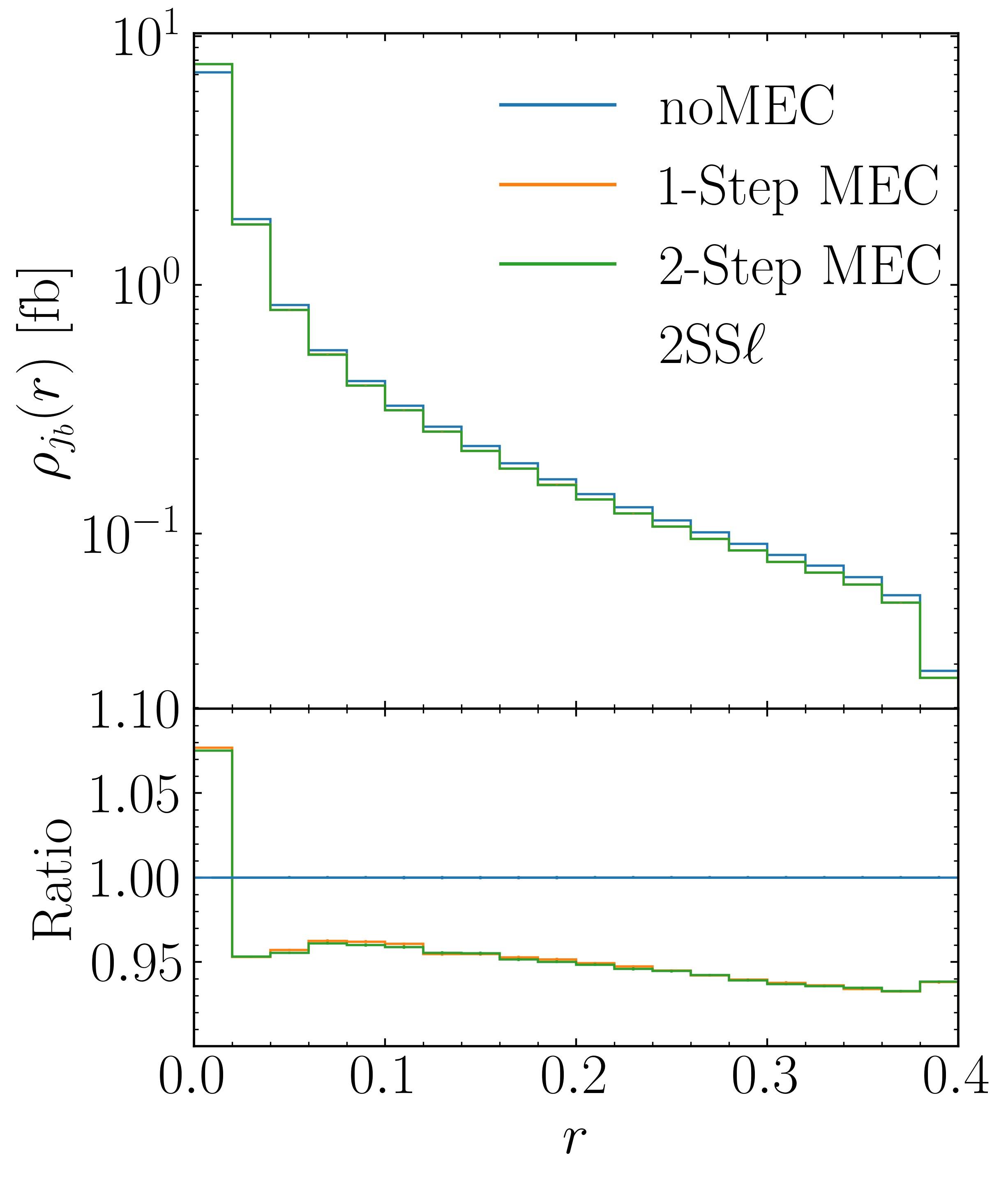}
\includegraphics[scale=0.49]{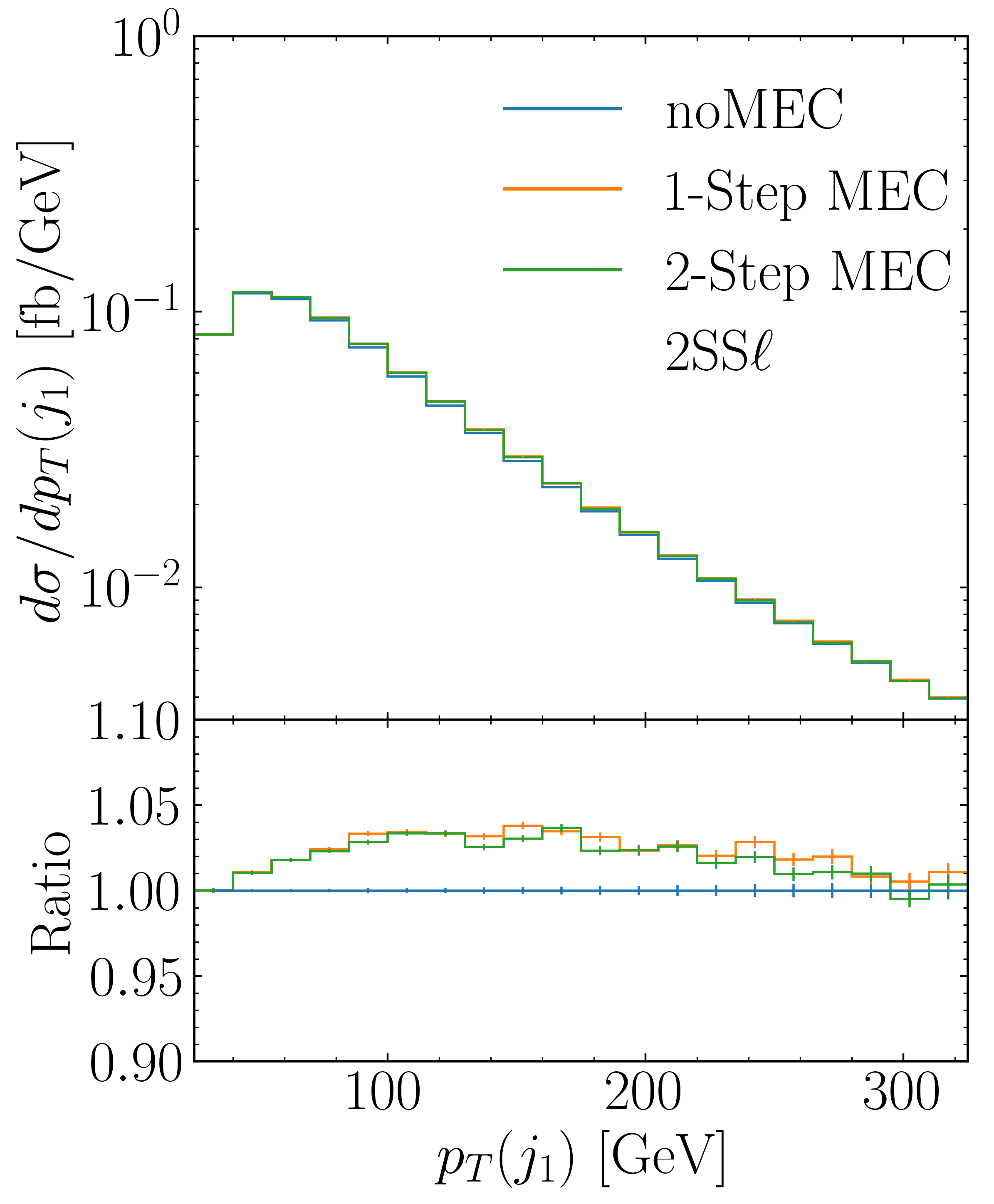}
\caption{The left plot shows the differential jet shape distribution of all $b$-jets  in the $2\rm{SS}\ell$ final state. The right plot shows the transverse momentum of the hardest light jet in the $2\rm{SS}\ell$ final state. The upper panel contains absolute predictions for the setup without MECs and two setups with MECs included. The lower panel contains the ratio of all predictions to noMEC.}
\label{fig4}
\end{figure}
In this section we present our predictions for $t\overline{t}W^{\pm}$ in the $3\ell$ and $2\rm{SS}\ell$ fiducial regions, matched to parton showers at NLO in QCD accuracy. For a comprehensive analysis, we directly compare predictions with MECs, generated via the 1-step and 2-step approaches discussed in section  \ref{explain_MEC}, to predictions without MECs. All three setups are presented for the default scale settings and fiducial cuts as defined in section \ref{setup}.\\
The integrated cross sections for the $2\rm{SS}\ell$ and $3\ell$ regions are:
\begin{eqnarray}
\rm{no~MEC} \qquad\qquad  \sigma^{t\overline{t}W^{\pm}}_{2\rm{SS}\ell} =  12.84~\rm{fb} \qquad\qquad \sigma^{t\overline{t}W^{\pm}}_{3\ell} = 3.92~\rm{fb} ~~\textcolor{white}{.}\\
\rm{with~MEC} \qquad\qquad  \sigma^{t\overline{t}W^{\pm}}_{2\rm{SS}\ell} =  13.12~\rm{fb} \qquad\qquad \sigma^{t\overline{t}W^{\pm}}_{3\ell} =  4.02~\rm{fb}~~.
\end{eqnarray}
The value we report for the the ``with MEC" case is valid for both the 1-step and 2-step approaches, since they yield identical cross sections within integration uncertainties. MECs induce a moderate increase of $2\%-3\%$ with respect to the ``no MEC" case. This is as expected, considering that they should not lead to large corrections for sufficiently inclusive observables.\\
To investigate whether significant MEC corrections arise, we are thus led to study differential distributions. The plot format consists of two panels: the upper panel contains the absolute predictions for all three setups and the lower panel shows the ratio to the setup where no MECs are applied. The final state signature is indicated in the legend for each plot. We also show integration uncertainties for each bin.\\
In fig.\ \ref{fig1} we show the transverse momenta of the charged leptons for the $2\rm{SS}\ell$ and $3\ell$ signatures. The 1-step and 2-step approaches are in good agreement for both observables. In the case of the hardest same-sign lepton for the $2\rm{SS}\ell$ signature, MECs induce a constant upwards shift of $2\%-3\%$. This is the case for all the $p_T$'s of the hardest and second-hardest same-sign leptons in the $2\rm{SS}\ell$  and $3\ell$ signatures. Since MECs are directly related to the QCD part of the top decay, the lepton kinematics are not directly sensitive to them.  Yet, through momentum shifts due to recoil, one could expect to see effects in some cases. In the case of the opposite-sign lepton, which stems exclusively from one of the top-quarks, we see a slight enhancement by MECs, of up to up to $5\%$, above $100~\rm{GeV}$. This effect is diluted in the case of the same-sign leptons, due to the lepton from the associated $W^{\pm}$-gauge boson.\\
Observables where we anticipate to see significant corrections are those related to jet kinematics. In fig.\ \ref{fig2} we show the number of all jets for the $2\rm{SS}\ell$  signature on the left, and for the $3\ell$ signature on the right. Within the plotted range between 2 and 7 jets, MECs induce shape differences by increasing the cross section for low multiplicities and decreasing it towards high multiplicities. This trend follows at different paces depending on the final state. This is due to the opposite-sign $W$-gauge boson decaying into partons $W\rightarrow q \overline{q}'$ for $2\rm{SS}\ell$. For the $2\rm{SS}\ell$ signature, MECs are positive and at $+4\%$ relative to noMEC predictions for $N_j \leq 4$. They drop to zero for $N_j=5$, and are negative, around $-3\%$, towards higher numbers of jets. On the other hand, for the $3\ell$ signature corrections are $+7\%$ in the first bin, and they vanish earlier, already at $N_j=3$. For higher jet numbers, MEC effects tend to a constant $-3\%$ decrease in the cross section. The quoted impact of MECs is similar for the 1-step and 2-step approaches, but there is a percent level difference for relevant bins in $N_j$, that is not covered by the integration error. The disagreement stems from the different treatment of QED radiation for the first radiation of $\mathbb{S}$-type events. In the 2-step approach, QED emissions are explicitly prohibited for the first emission of $\mathbb{S}$-type events, whereas in the full \pythia{} shower there is no such restriction.\\
Next, in fig.\ \ref{fig3}, we look at $b$-jet kinematics for the $2\rm{SS}\ell$ fiducial region, specifically the transverse momenta of the first and second hardest $b$-jets. We limit ourselves to showing and discussing results for this final state, since the $3\ell$ case is comparable. For $b$-jets produced close to the phase space cut of $p_T \gtrsim 25~\rm{GeV}$ threshold, the cross section is reduced by $-6\%$ for $p_T(b_1)$ and $-3\%$ for $p_T(b_2)$. For $p_T$'s above $50~\rm{GeV}$, the corrections become positive and increase the cross section by around $3\%$-$6\%$. Above $300~\rm{GeV}$ MECs diminish to around $+1\%$ or less for $p_T(b_1)$. We do not plot this range for $p_T(b_2)$ due to small statistics. In this case too, we find that the 1-step and 2-step methods are indistinguishable.\\ 
Equipped with this information, we can now attempt to shed some light on how MECs affect the kinematics of the process. Earlier we saw that MECs lead to an increase in the cross section for $N_j=2$, which are exactly the two $b$-jets required by our cuts. Now, we saw that MECs increase the probability of encountering harder $b$-jets in the $p_T$ range $[50,300]~\rm{GeV}$. These findings point to the explanation that the extra radiation stemming from MECs is recombined with the original $b$-parton more often, relative to emissions from the shower.\\
To investigate the $b$-jets in further detail, we can look at the differential jet shape distribution, which is defined as in ref.\ \cite{CMS:2012oyn}
\begin{eqnarray}
\rho(r) = \frac{1}{\Delta r}  \frac{1}{N_{\rm{jets}}} \sum_{\rm{jets}} \frac{p_{\perp}(r-\Delta r/2,r+\Delta r/2)}{p_{\perp}(0,R)} ~~.
\end{eqnarray}
It characterizes the transverse momentum  of all jet constituents that are inside a jet-annulus $[r-\Delta r/2,r+\Delta r/2]$. Here $\Delta r$ denotes the width of the annulus, as well as the bin-width. The quantity $r$ is defined as $r=\sqrt{(\Delta y)^2+(\Delta \phi)^2}$ with respect to the jet momentum for each parton inside the jet. This observable is defined to be quite sensitive to the distribution of the internal momenta inside the jet.\\
On the left of fig.\ \ref{fig4} we show the differential jet shape distribution for $b$-jets in the $2\rm{SS}\ell$  final state. MECs induce a distinct enhancement of $+7\%$ in the first bin, which can be due to collinear MEC radiation being recombined with the original $b$-jet more often or an increased chance that the $b$-parton carries more of its momentum when MECs are turned on. For the rest of the distribution MECs diminish the cross section by $5\%$-$7\%$. In the case of $3\ell$ we find similar results.\\
In addition, we can also analyze the kinematics of light jets. On the right of fig.\ \ref{fig4} we show the transverse momentum of the hardest light jet in the $2\rm{SS}\ell$ final state. To this end, we add an the additional requirement of a light jet in the final state, which fulfills the same phase space cuts as the $b$-jets. We notice that there is an increase of at most $+3\%$ if MECs are included in the calculation of the cross section. The enhancement is present in the range $p_T \in [50,250]~\rm{GeV}$, and it is not constant over the plotted region. This shape difference is similar to the one of the opposite-sign lepton. The hardest light jet for the $2\rm{SS}\ell$ should predominantly originate from one of the top-quarks.\\
Finally, we have also investigated different options in the simple shower of \pythia. We find that enabling further MECs in the shower using \texttt{MEafterFirst}, does not lead to any significant differences. Another interesting feature, that is relevant for the $t\overline{t}W^{\pm}$ process, is the choice for the resonance-final dipole \texttt{TimeShower:recoilStrategyRF} \cite{Bierlich:2022pfr,Brooks:2019xso}. These kinds of dipoles appear due to successive shower emissions in the decay products of a colored resonance, such as the top-quark. In such a case, if a final state-parton that has the color of the top radiates, the current default setting in \pythia{} (\texttt{option 1}) is that the $b$-particle takes the recoil. Opting for some of the alternative options (\texttt{option 2,3}), allows for the $W$-gauge boson to take on the role of the recoiler instead, with an additional eikonal correction factor. Opting for the either \texttt{option 2} or \texttt{option 3} leads to a similar impact to that of MECs in the integrated cross section and distributions of the noMEC predictions. If MECs are turned on, the impact of employing these alternative options is rather small by comparison.  
%
%
%
%
%
%
\section{Summary and Conclusions}
\label{finish}
In this work, we presented a prescription that allows for enabling MECs in the simple shower of \pythia, while remaining consistent with the \mcatnlo{} matching procedure. This prescription was implemented and employed to calculate predictions for the  $t\overline{t}W^{\pm}$ process in the $2\rm{SS}\ell$ and $3\ell$ final states.\\
MECs are an interesting feature in \pythia{} that offer a reliable and computationally efficient way with the potential to improve current automated predictions. They can upgrade the accuracy of top quarks decays to approximately NLO QCD in the shower, alongside matching and jet merging procedures. The prescription we propose, is based on the idea that the \mcatnlo{} parton shower matching can be executed in a single phase, and then separated from the rest of the shower evolution. This eliminates the double-counting problem. We note, that this separation is independent of the particular process. Once the matching has been taken care of, one can e.g.\ turn on MECs without encountering any double-counting issues. We have validated our approach and compared directly to the suggestion from ref.\ \cite{Frixione:2023hwz}. Broadly we find good agreement, up to distributions such as $N_j$. In this case the 1-step and 2-step approaches differ by about one percent. The source of the difference is that QED emissions are not allowed for the first emissions of $\mathbb{S}$-events for the 2-step approach.\\
Another aspect of the $2$-step method, is that it opens up the possibility of using advanced shower types in the second phase. Available options in \pythia{} include Dire \cite{Hoche:2015sya} and Vincia \cite{Giele:2007di,Giele:2011cb}.\\
A main focus of our study has been the impact of MECs in the decay of the top-quark for the $t\overline{t}W^{\pm}$ process. We generated predictions at NLO in QCD matched to parton showers with and without MECs included. The corrections contribute positively, by $2\%$-$3\%$, to the integrated fiducial cross section. It is observables that are related to the kinematics of the top decay products that receive larger MEC contributions in certain regions of the phase space. In the case of $b$-jets, MECs shift transverse momenta close to the phase-space cut threshold, towards harder $p_T$'s above $50~\rm{GeV}$. This pattern can be explained by the additional radiation induced by MECs re-combining with the original $b$-jets more frequently than other shower emissions. This effect is also visible in the distribution over the number of all-jets $N_j$. It can contribute to the cross section by up to $\pm 6\%$ in sensitive regions for these observables. MECs can also indirectly impact the kinematics of the other top decay products, the $W$-gauge bosons, through the recoil mechanism. This can be clearly seen in the kinematics of the opposite-sign lepton for $3\ell$ and the hardest light jet for $2\rm{SS}\ell$. In a recent study in ref.\ \cite{ATLAS:2023gon}, ATLAS investigates related distributions in a different fiducial region compared to ours and finds that all theoretical predictions they consider undershoot measurements for $N_j$. This gap increases for the $2\rm{SS}\ell$ final state towards fewer numbers of jets. We saw that MECs contribute positively to the cross section in this particular phase space region. Thus, they might help in reducing the gap by some amount in sensitive phase space regions, but they cannot account for the significant differences present between theoretical predications and experimental measurements.\\
Given their impact, particularly in jet related observables, we suggest that MECs should be considered when generating predictions with \mcatnlo{} matching in \pythia. 
%
%
%
%
%
%
\section*{Acknowledgments}
This work is done in the context of and supported by the Swedish Research Council contract numbers 2016-05996 and 2020-04423. J.N. thanks Torbjörn Sjöstrand for helpful discussions and the Galileo Galilei Institute for Theoretical Physics for the hospitality and the INFN for partial support during the completion of this work.
%
%
%
%
%
%
\bibliography{References} 
\bibliographystyle{JHEP}
\end{document}